\documentclass[%
 reprint,
 twocolumn,
 amsmath,amssymb,
 aps,
 prl,
 superscriptaddress,
]{revtex4-2}

\usepackage{amsmath,hyperref}
\usepackage{amssymb}
\usepackage{braket}
\usepackage{graphicx}
\usepackage{dcolumn}
\usepackage{bm}
\usepackage{xcolor}
\usepackage{orcidlink}
\newcommand{\tr}{\mathrm{Tr}}

\newcommand{\hatH}{\hat{H}}
\newcommand{\eps}{\boldsymbol{\epsilon}}

\begin{document}

\title{Predicting open quantum dynamics with data-informed quantum-classical dynamics }

\author{Pinchen Xie\orcidlink{0000-0002-9330-4032}}
\thanks{pinchenxie@lbl.gov}
\affiliation{Applied Mathematics and Computational Research Division, Lawrence Berkeley National Laboratory, Berkeley, CA 94720, USA}
\author{Ke Wang\orcidlink{0009-0001-3225-887X}}
\affiliation{Department of Mathematics, Pennsylvania State University, University Park, PA 16802, USA}

\author{Anupam Mitra\orcidlink{0009-0004-6711-2477}}

\affiliation{Applied Mathematics and Computational Research Division, Lawrence Berkeley National Laboratory, Berkeley, CA 94720, USA}

\author{Yuanran Zhu \orcidlink{0000-0001-6851-4161}}
\affiliation{Applied Mathematics and Computational Research Division, Lawrence Berkeley National Laboratory, Berkeley, CA 94720, USA}

\author{Xiantao Li\orcidlink{0000-0002-9760-7292}}
\affiliation{Department of Mathematics, Pennsylvania State University, University Park, PA 16802, USA}

\author{Wibe Albert de Jong\orcidlink{0000-0002-7114-8315}}
\affiliation{Applied Mathematics and Computational Research Division, Lawrence Berkeley National Laboratory, Berkeley, CA 94720, USA}


\author{Chao Yang \orcidlink{0000-0001-7172-7539}}
\thanks{cyang@lbl.gov}
\affiliation{Applied Mathematics and Computational Research Division, Lawrence Berkeley National Laboratory, Berkeley, CA 94720, USA}

\date{\today}

\begin{abstract}
We introduce a data-informed quantum-classical dynamics  (DIQCD) approach for predicting the evolution of an open quantum system. The equation of motion in DIQCD is a Lindblad equation with a flexible, time-dependent Hamiltonian that can be optimized to fit sparse and noisy data from local observations of an extensive open quantum system. 
We demonstrate the accuracy and efficiency of DIQCD for both experimental and simulated quantum devices. We show that DIQCD can predict entanglement dynamics of ultracold molecules (Calcium Fluoride) in optical tweezer arrays. DIQCD also successfully predicts carrier mobility in organic semiconductors (Rubrene) with accuracy comparable to nearly exact numerical methods.

\end{abstract}

\maketitle


Modelling a quantum device in a noisy environment can be simplified by focusing on a specific subsystem and treating the remaining degrees of freedom as an effective bath that interacts with the subsystem. There have been various methods that model equations of motion (EOM) of such an open quantum system with a trade-off between accuracy and efficiency~\cite{lindblad1976generators,gorini1976completely,garg1985effect,tanimura1989time,garraway1997decay,diosi1997non,tanimura2006stochastic,xu2007dynamics,suess2014hierarchy, tamascelli2018nonperturbative, wang2024simulation}. 
On the high-accuracy end, pseudomode methods~\cite{garraway1997decay,tamascelli2018nonperturbative,li2021markovian,park2024quasi,park2024quasi, huang2025coupled} expand the system's Hilbert space with effective modes that encode the bath's correlation function. On the high-efficiency end, the plain Lindblad equation sketches a dissipative environment without introducing auxiliary degrees of freedom.
In between, mixed quantum-classical dynamics (MQCD) may strike a balance. With additional classical degrees of freedom, MQCD can describe coherent fluctuations that are otherwise reduced to mean-field dissipators in the plain Lindblad equation.
However, phenomenological MQCD approaches such as Ehrenfest dynamics~\cite{ehrenfest1927bemerkung} and surface hopping~\cite{tully1990molecular} can suffer from unsystematic error control relative to exact quantum dynamics~\cite{subotnik2016understanding}, hindering  accurate predictions of material properties or device metrics. 
Furthermore, for systems where the environment does not have a particle-based semiclassical representation, phenomenological MQCD becomes irrelevant. For example, one needs to devise special-purpose MQCD models for cold-atom systems~\cite{shaw2024learning}.  
  
We introduce Data-Informed Quantum-Classical Dynamics (DIQCD), a method that places MQCD in a general context by introducing variational capacity to the EOM.  We have developed a framework to optimize a general MQCD EOM, parameterized as a time-dependent Lindblad equation, using time-series data of quantum observable measurements. DIQCD is practical because it requires measurement data from the quantum subsystem rather than the environment. DIQCD is suited for sparse and noisy data due to its restricted variational flexibility, which prevents overfitting.

We demonstrate DIQCD's applications with two distinct case studies. The first is on dynamics of Calcium-fluoride (CaF) molecular qubits~\cite{holland2023demand}. We train DIQCD on sparse experimental data from a single CaF molecule and then predict two-qubit entanglement dynamics. The second is on quantum transport in Rubrene crystal, where DIQCD, trained on data from simulating a single Rubrene molecule, predicts carrier mobility with an accuracy similar to a nearly exact time-dependent density matrix renormalization group (TD-DMRG) simulation~\cite{li2020finite} but at a lower computational cost, bringing a new perspective to modeling band-like transport.

{\bf Methods} -- 
In the following, we call the quantum subsystem of interest the ``system'', and the rest of the degrees of freedom the ``environment''. The DIQCD approach is made up of a flexible EOM of the system, a loss function for optimizing the EOM, and the relevant training and simulation algorithms. 
The EOM is a Lindblad equation with a time-dependent Hamiltonian $\hat{H}_{\eps}(t)$:
\begin{equation}\label{lindblad}
\small
    \frac{d \hat{\rho}_{\eps}(t) }{\partial t} = -i[\hat{H}_{\eps}(t),\hat{\rho}_{\eps}(t)] + \sum_k \gamma_k (\hat{L}_k\hat{\rho}_{\eps}(t) \hat{L}_k^\dagger-\frac{1}{2}\{\hat{L}_k\hat{L}_k^\dagger, \hat{\rho}_{\eps}(t)\}).
\end{equation}
$\hat{\rho}_{\eps}(t)$ is the density matrix of the system. $\{ \hat{L}_k \}$ is a finite set of static jump operators. $\hat{H}_{\eps}(t)=\hat{H}_0 + \hat{H}_c(t) + \sum_{j=1}^M f_j(\eps(t)) \hat{S}_j$ combines the static system Hamiltonian $H_0$, the external control $\hat{H}_c(t)$, and perturbing Hermitian operators $\hat{S}_j$.  $f_j(\eps(t))$ is a scalar function of the multidimensional, classical dynamical processes $\eps(t)$ that encodes information about the environment. 
As a special case, Eq.~\eqref{lindblad} becomes Ehrenfest dynamics when $\eps(t)$ represents atomic nuclei evolving under the average potential energy surface $\tr(\hat{H}_{\eps}(t) \rho_{\eps})$. In DIQCD, we allow $\eps(t)$ to be generic dynamical processes that can be integrated by an explicit Markovian integrator with or without auxiliary degrees of freedom $\boldsymbol{\xi}(t)$. This means $\eps(t+\delta t)$ and $\boldsymbol{\xi}(t+\delta t)$ are determined by $\eps(t)$ and $\boldsymbol{\xi}(t)$ in an integration scheme with $\delta t$ as the time step. These requirements are imposed such that $\eps(t)$ can evolve concurrently with the structure-preseving integration~\cite{cao2021structure} of the Lindblad equation. And eventually, the parameters controlling the EOM of $\eps(t)$ can be optimized together with other system parameters (such as $\gamma_k$) through chain rules, such that the behavior of $\hat{\rho}_{\eps}(t)$ can match time-series data of the system. A Langevin process $\epsilon_0(t)$ is an example of a flexible classical process. The evolution of $\epsilon_0(t)$ can be approximated by ~\cite{zhang2019unified} $\epsilon_0(t+\delta t) = e^{- \delta t/\tau}\epsilon_0(t) + A\sqrt{1 - e^{-2\delta t/\tau}}\xi_0(t)$ for some parameters $\tau$ and $A$. $\xi_0(t)$ is a Gaussian white noise. Simpler examples of flexible processes include periodic signals and Gaussian white noise. Complex examples include molecular dynamics on a parameterized potential energy surface.

In DIQCD, the expectation of any system observable $\hat{O}$ at time $t$ is given as $O(t) = \langle O_{\eps}(t) \rangle_{\eps}=\langle \tr (\hat{O}\hat{\rho}_{\eps}(t))\rangle_{\eps}$, where $\langle \cdot \rangle_{\eps}$ represents the average over realizations of $\eps(t)$. 
DIQCD is trained on time-series data at discrete time points $\{t_1, t_2, \cdots \}$ with the mean-squared loss $\mathcal{L}=\sum_{ij} (O_i(t_j) - O_i^*(t_j))^2$. The data $O_i^*(t)$ comes from measurements performed on either an experimental or simulated system. For example, for an atom-based quantum computing platform,  $\hat{O_i}$ can be a Pauli operator acting on the $i$-th qubit, and $\eps(t)$ conceptually encapsulates dynamical information about optical traps, environmental radiation, etc. In DIQCD, one does not need to model $\eps(t)$ with precise knowledge of the environment. Various classical processes can be incorporated. A guiding principle for practical modeling is to employ a collection of processes characterized by flexible time scales~\footnote{such as the damping time in a Langevin process and the angular frequency of an harmonic oscilator}, initialized to broadly encompass the range of time scales from $\delta t$ up to the quantum coherence time, and extending to infinity to represent shot-by-shot noise. 
Other than that, DIQCD does not specify the choice of $\{\hat{S}_j\}$ or the choice of system degrees of freedom. These considerations should be made with a phenomenological understanding of a quantum device. 

The DIQCD model is trained using a forward-backpropagation loop, in which $\rho_{\eps}(t)$ and $\eps(t)$ are propagated from $t=0$ to $T$ via  straightforward numerical integration of Eq.~\eqref{lindblad} and the associated classical EOMs. We calculate the loss function $\mathcal{L}$, backpropagate the loss over the entire trajectory of $\rho_{\eps}(t)$ and $\eps(t)$, and update all flexible parameters with accumulated gradients. 
Details on the implementation of DIQCD, i.e., the Python package QEpsilon~\cite{qepsilon}, are reported in Appendices. 

\begin{figure*}[tb]
    \centering
    \includegraphics[width=\linewidth]{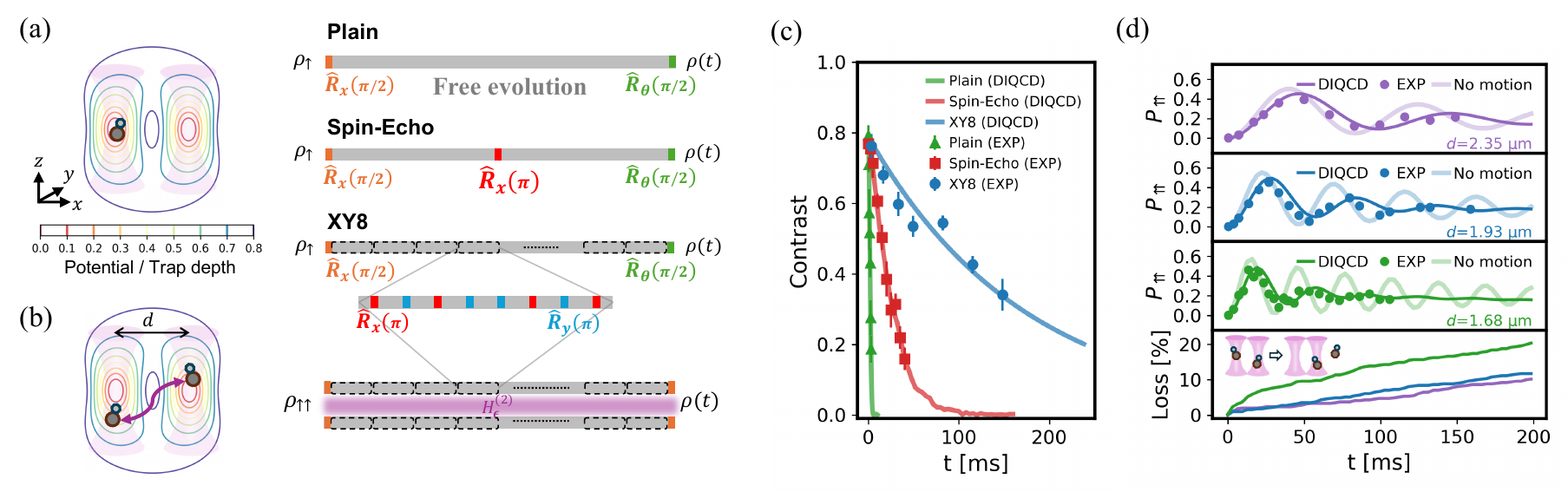}
    \caption{ (a) Optically trapped CaF molecule and three control schemes used in Ramsey experiments. Contour lines represent the equipotential surface of the trap, with energy levels indicated in the color map.   (b) A pair of CaF molecules and the control scheme for Bell state generation. (c) The contrast $C(t)$ (scaled by $\varsigma$) as a function of circuit duration $t$. Circle, square, and triangle markers represent $C^{\text{EXP}}(t)$ data.  (d) The top three panels show $P_{\uparrow\uparrow}$ (scaled by $\varsigma^2$) as a function of $t$. The bottom panel shows the rate of qubit loss as a function of $t$.}
    \label{fig:caf}
\end{figure*}

{\bf CaF molecular qubits} -- Reconfigurable optical tweezer arrays of ultracold molecules have emerged as a promising platform for precision measurement and quantum simulation~\cite{Kaufman2021TweezerReview, Doyle2021CaFCoherence,holland2023demand, Doyle2023SpinExchange}. A precise {\it ab initio} model to capture multi-scale quantum dynamics of this platform is not yet available. DIQCD can circumvent this difficulty and establish a reliable and scalable dynamical model based on data collected from a minimal unit of the platform: a single trapped molecule.

For example, we consider a recent realization of the platform~\cite{holland2023demand} (see Appendices for details), where two CaF molecules are trapped in respective optical tweezers separated by a distance of $d$. The trapping potential is illustrated in Fig.~\ref{fig:caf}(a-b). Let $\mathbf{R}(t)=(\mathbf{r}_1(t), \mathbf{r}_2(t))$ be the center-of-mass positions of the two molecules~\footnote{For the experimental molecular temperature~\cite{holland2023demand} without further cooling~\cite{lu2024raman}, quantum fluctuation in $\mathbf{r}_1$ and $\mathbf{r}_2$ is insignificant.}. The quantum system here is relevant to two hyperfine states $\ket{\uparrow}$ and $\ket{\downarrow}$ of each molecule.
Let $\hat{S}_i^{\alpha}=\frac{\hbar}{2}\hat{\sigma}_i^{\alpha}$ ($i=1,2$; $\alpha=x,y,z$) be spin-1/2 operator for molecule-$i$. 
An effective Hamiltonian $H_{\eps}(t)$, used by DIQCD to describe the system, contains the one-body terms
\begin{equation}
\small
H^{(1)}_{\eps}(t) = \sum\nolimits_{j=1}^L \epsilon_{1,j}(t)\hat{S}_1^z + \sum\nolimits_{j=1}^L \epsilon_{2,j}(t) \hat{S}_2^z,
\end{equation}
and the two-body dipole-dipole interaction
\begin{equation}
\small
H^{(2)}_{\eps}(t) =  J(\mathbf{r}_1, \mathbf{r}_2) ( \hat{S}_1^x \hat{S}_2^x + \hat{S}_1^y\hat{S}_2^y).
\end{equation}
Here, $J(\mathbf{r}_1, \mathbf{r}_2)= \frac{J_0}{r_{12}^3} (1-3\cos^2\theta')$ with 
$J_0$ being the coupling constant, $r_{12}$ being $\|\mathbf{r}_1 - \mathbf{r}_2\|$,  $\theta'$ being the angle between $(\mathbf{r}_1 - \mathbf{r}_2)$ and the quantization axis ($y$-axis).  

From phenomenological understanding of the environments, we take an ansatz with $L=6$: $\epsilon_{i,l}$ ($i=1,2$; $l=1,2,3,4$) are periodic processes with fixed frequency $\omega_l = l\times60\text{Hz}$ and flexible~\footnote{The word ``flexible'' indicates quantities that will be optimized with data.} amplitudes. $\epsilon_{i,5}$ are flexible, overdamped Langevin processes. $\epsilon_{i,6}$ are time-independent, shot-by-shot noises with uniformly distributed initial values sampled from a flexible interval. Here,  $\epsilon_{i,l}$ ($l=1,2,3,4$) represent line noises from power sources. $\epsilon_{i,6}$ represents noises with a time scale above ms. The Langevin process $\epsilon_{i,5}$ with a continuous spectrum approximates the mix of other sources of noise with time scales comparable to execution time. The choices of these processes are not unique, but should be organized in a way that adaptively covers a large range of time scales encompassing the time scale of system dynamics. 

The DIQCD EOM is then given by Eq.~\eqref{lindblad} with one-body jump operators $\{\hat{S}_i^{x}, \hat{S}_i^{z}\}_{i=1,2}$ and two flexible damping rates $\gamma_x$ and $\gamma_z$, as a minimal parameterization of one-body dissipation. The EOM of $(\mathbf{r}_1, \mathbf{r}_2)$ is standard isothermal molecular dynamics (see Appendices) without non-adiabatic force, because  $\tr(\rho H^{(2)}_{\eps})$ is negligible for the all values of $d$ used in experiments. 

We train the DIQCD model on data from a series of one-molecule Ramsey experiments~\cite{holland2023demand}, where the motion of the molecule's position is irrelevant. Experiments were performed with initial state  $\ket{\uparrow}$ and three external control schemes as illustrated in Fig.~\ref{fig:caf}(a): a plain scheme, a spin-echo scheme, and an XY8 scheme. These schemes probe the environment over different time scales~\cite{viola1999dynamical}, allowing DIQCD to distinguish noise components across these scales.
Since the external controls consist of short pulses, we do not model them as a continuous control $\hat{H}_c(t)$, but as instantaneous one-body unitary rotations $\hat{U}=\hat{R}_{\hat{n}}(\cdot)$ along axis $\hat{n}$, followed by an instantaneous quantum error channel. Their combined effect is given as 
 $\rho(t+\delta t) = \sum_{\alpha=1}^4 \hat{K}_\alpha \hat{U}\rho(t) \hat{U}^\dagger \hat{K}_\alpha^\dagger$, where the Kraus operators $\hat{K}_\alpha$ represent a depolarization error channel~\footnote{$\hat{K}_0=\sqrt{1-p}\hat{I}$, $\hat{K}_{1}=\sqrt{\frac{p}{3}}\hat{\sigma}_x$, $\hat{K}_{2}=\sqrt{\frac{p}{3}}\hat{\sigma}_y$, $\hat{K}_{3}=\sqrt{\frac{p}{3}}\hat{\sigma}_z$} controlled by a flexible scalar $p$. 
The collected data from the Ramsey experiments is the spin-up probability $P_{\uparrow}(t,\theta)$ as a function of circuit duration $t$ and the angle $\theta$ associated with the last control pulse (see Fig.~\ref{fig:caf}(a)). Let $C(t)=|P_{\uparrow}(t,\pi) - P_{\uparrow}(t,0)|$. Experiments yield 24 $C^{\text{EXP}}(t)$ data points shown in Fig.~\ref{fig:caf}(c), with a state preparation fidelity $\varsigma\approx 0.79$. 
The parameters embedded in $\{\epsilon_{1,j}\}\cup\{\epsilon_{2,j}\}\cup\{\gamma_x, \gamma_z, p\}$ are optimized by matching $C(t)$ predicted by a single DIQCD model, but with three control schemes, with all 24 $C^{\text{EXP}}(t)/\varsigma$ data points. The solid lines in Fig.~\ref{fig:caf}(c) are $\varsigma C(t)$ from the optimized DIQCD, showing excellent agreement with experiments across multiple time scales. Such success would not be achieved by the plain Lindblad equation. All parameters in DIQCD converge to substantially non-zero values~\footnote{See Supplementary Code in~\cite{qepsilon} for evolution of all parameters during the training.}, indicating a separation of noise over frequency regime. The optimized magnitude of the shot-by-shot noise ($\epsilon_{i,6}$) is comparable to other sources of decoherence, suggesting that low-frequency noises may be further disentangled if one can perform experiments probing a larger time scale. 


Next, we model two-molecule system by including the thermal dynamics of $\mathbf{R}(t)$ and the dipole-dipole interaction $H^{(2)}_{\eps}(t)$ in DIQCD. Radial and axial temperature of the molecules are set to $T_r=6\mu K$ and $T_a=18\mu K$ (from Raman Thermometry~\cite{lu2024raman}) for isothermal simulation of $\mathbf{R}(t)$.  The resultant DIQCD model is not trained on any two-molecule data. We will validate its prediction against experimental results on Bell state creation (See Fig.~\ref{fig:caf}(b) and Appendices). The ideal (without decoherence or molecular motion) probability of observing $\ket{\uparrow\uparrow}$ from final measurement is $P_{\uparrow\uparrow} = \sin^2(\frac{J_0t}{4\hbar d^3})$. The oscillation in $P_{\uparrow\uparrow}$ is damped in experimental measurements $P^{\text{EXP}}_{\uparrow\uparrow}$, plotted in Fig.\ref{fig:caf}(d) for different tweezer separation $d$. DIQCD can predict $\varsigma^2P_{\uparrow\uparrow} $ (solid lines in top three panels of Fig.\ref{fig:caf}(d)) in excellent consistency with $P^{\text{EXP}}_{\uparrow\uparrow}$. Moreover, molecular dynamics predict the rate of qubit loss \footnote{The qubit loss is severe for $d<1.68 \mu m$ due to parametric heating of molecules. For the same reason, the current DIQCD model with a fixed molecular temperature can not be directly applied to $d<1.68 \mu m$.} (see Fig.\ref{fig:caf}(d)) due to molecules hopping from one optical trap to another. The importance of molecular dynamics is revealed when we simulate DIQCD with molecules fixed to tweezer centers --- the resultant $\varsigma^2P_{\uparrow\uparrow} $ (semi-transparent lines in Fig.\ref{fig:caf}(d)) shows underestimated damping and overestimated frequency of oscillation. 

One can generalize the DIQCD model to a larger array of tweezers with pairwise dipolar interactions, or extend the approach to quantum devices involving instantaneous feedback action~\cite{wiseman1993quantum}. Furthermore, DIQCD can model molecules cooled to almost the motional ground state by replacing classical molecular dynamics with ring-polymer path-integral molecular dynamics~\cite{marx1996ab}.


\begin{figure}[tb]
    \centering
    \includegraphics[width=0.9\linewidth]{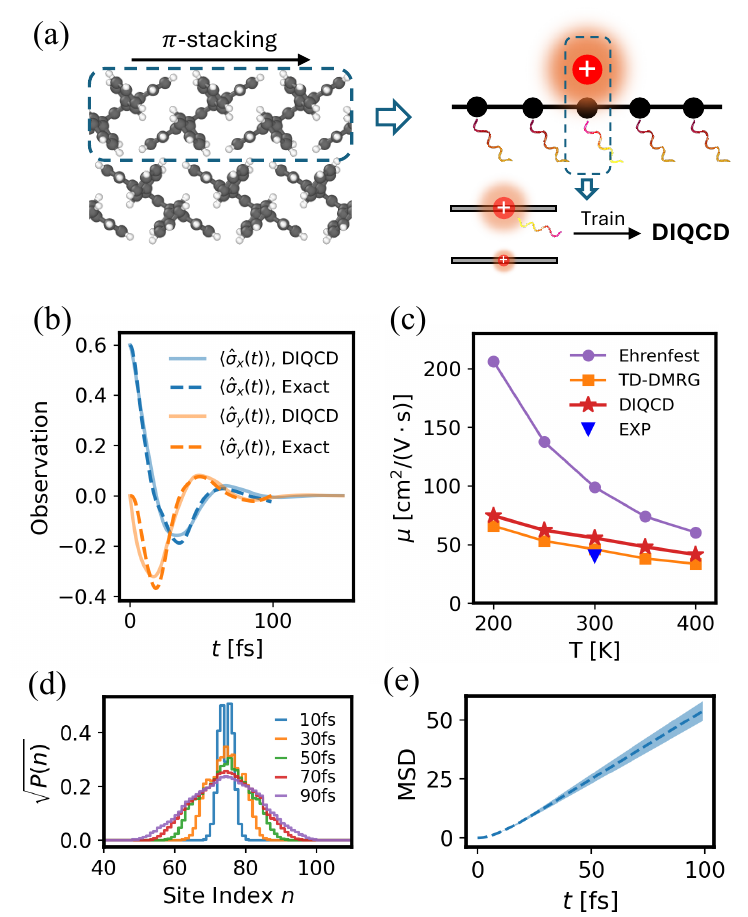}
    \caption{ (a) Rubrene crystal (left) and the effective model for carrier transport (right). (b) Comparison of data and DIQCD on average $\langle \hat{\sigma}_x(t) \rangle$ and $\langle \hat{\sigma}_y(t)\rangle$ for $T=300$K. (c) Carrier mobility as a function of temperature. The blue triangle is an experimental estimation of the intrinsic mobility~\cite{takeya2007very}.  (d) The square root of site occupation on the $L=150$ molecular lattice. Snapshots are obtained from a DIQCD simulation lasting 100fs for $T=300$K. (e) The mean squared displacement (MSD=$\langle \tr(\rho_{\boldsymbol{\epsilon}}(t)n^2) -\tr^2(\rho_{\boldsymbol{\epsilon}}(t) n )\rangle_{\boldsymbol{\epsilon}}$) as a function of time  for $T=300$K. The shades represent standard deviation. }
    \label{fig:rubrene}
\end{figure}

{\bf Rubrene Crystal} -- 
The development of flexible electronics requires screening of high-mobility organic semiconductors~\cite{mei2013integrated}. 
Here we show the relevance of DIQCD with first principles-based prediction of the intrinsic carrier mobility ($\mu$) of Rubrene crystal, which has an excellent hole mobility~\cite{takeya2007very, ren2017negative}. The high mobility is due to the coherent quantum transport predominantly along the $\pi$-stacking direction. The effective model for transport in Rubrene crystal can therefore be reduced to a 1D Holstein model with the Hamiltonian $\hatH = \hatH_{\text{e}} + \hatH_{\text{ph}} + \hatH_{\text{ep}}$. 
Let $n$ be the label of a Rubrene molecule on the 1D molecular lattice along the stacking direction. Let $\hbar=1$. The single-carrier tight-binding Hamiltonian $\hatH_{\text{e}} = V\sum_n (c_{n+1}^\dagger c_n + c_n^\dagger c_{n+1} )$ describes intermolecule carrier hopping. The harmonic Hamiltonian $\hatH_{\text{p}} = \sum_{n,m}  \omega_m b^\dagger_{n,m} b_{n,m}$ is associated with the bosonic vibrational modes of each molecule. The interaction $\hatH_{\text{ep}} = \sum_{n,m} g_m\omega_m (b^\dagger_{n,m} + b_{n,m}) c^\dagger_n c_n$ is associated with electron-phonon coupling, of which the strength is characterized by the polaron binding energy $\lambda=\sum_m g^2_m \omega_m$. 

Modeling quantum transport is challenging when $\lambda$ is of the same order of magnitude as $V$\footnote{This happens to be the case for most high-mobility organic semiconductors}, a regime where Marcus theory is inaccurate and traditional MQCD approaches offer results with uncontrolled error. While TD-DMRG can simulate unitary dynamics driven by $\hatH$ with nearly full accuracy, its application is restricted to small length scales, which allows the determination of $\mu$ with the Kubo formula, but not real-time, mesoscale transport \footnote{ The success of TD-DMRG is also limited to 1D due to the 1D nature of the matrix product state and the numerical inefficiency of higher-dimensional tensor networks. }.
In contrast, by treating the carrier as the system and bosonic modes as the environment, we demonstrate that DIQCD can predict $\mu$ with accuracy comparable to TD-DMRG, while offering superior numerical efficiency.  DIQCD requires training data from only a single Rubrene molecule, since the main mechanism slowing down the coherent transport considered here is phase decoherence from local system-environment interaction. 

We adopt the model of $\hat{H}$ with $V=83$meV, $\lambda=73$meV,  and nine renormalized vibration modes for each molecule with $g_m\in[0,1]$, and $\omega_m\in[84, 1594]\text{cm}^{-1}$ (see Appendices for details).
These parameters were obtained in Ref.~\cite{jiang2016nuclear, li2020finite} from a hybrid-level density functional theory ~\cite{Becke1993, Lee1988, g09}. Here, we are mainly interested in deriving $\mu$ near the room temperature. For $T=\text{200K, 250K, 300K, 350K and 400K}$, we generate one-molecule data by simulating the unitary evolution driven by an effective spin-boson Hamiltonian 
\begin{equation}
    \small
    \hat{H}^{\text{eff}}= \sum_{m=1}^9  \omega_m b^\dagger_{m} b_{m} + \sum_{m=1}^9 \frac{1}{2}g_m\omega_m (b^\dagger_{m} + b_{m}) (1+\hat{\sigma_z}),
\end{equation}
with the initial state $(\sqrt{\phi}\ket{\uparrow}+\sqrt{1-\phi}\ket{\downarrow})\otimes \ket{n_1, n_2, ..., n_9}$.  The Pauli operators ($\hat{\sigma}_{\alpha=x,y,z}$) are associated with the two states $\ket{\uparrow}$ and $\ket{\downarrow}$, representing respectively the presence and absence of a carrier. The bosonic state $\ket{n_1, n_2, ..., n_9}$ associated with nine vibrational modes is sampled from the Boltzmann distribution $\exp(-\beta\sum_{m=1}^9  \omega_m b^\dagger_{m} b_{m})$. The mean and standard deviation (concerning the sampled trajectories) of  $\langle\hat{\sigma}_x(t)\rangle$ and $\langle\hat{\sigma}_y(t)\rangle$ ($t\in[0,100]$fs) are obtained for each $T$ from a sufficient number of quantum trajectories initiated with the same $\phi$~\footnote{Results are not sensitive to the choice of $\phi$.}, but different pure states. 

These data are used to optimize a one-site DIQCD model with 
$\hat{H}^{\text{eff}}_{\eps}(t)=\epsilon_0 \hat{\sigma_z} +  \sum_{m=1}^9  \frac{1}{2} \epsilon_m(t) (1+\hat{\sigma_z})$ and the jumping operator $(1+\hat{\sigma_z})/2$ ($\gamma$ is the damping rate), for each $T$ separately. Here, $\epsilon_0$ and $\gamma$ are flexible, time-independent scalars. $\epsilon_m(t)$ is chosen to be a periodic sinusoidal signal with flexible amplitude, fixed frequency $\omega_m$, and random initial phases. These choices are motivated by the fact that, at high temperatures, the quantum statistics of phonons converge to classical statistics of harmonic oscillators. Traditional MQCD approaches ~\cite{wang2011mixed} adopt this rigid classical approximation also at low temperatures, leading to uncontrolled error from quantum kinetic energy. In contrast, DIQCD can incorporate excessive kinetic energy with a flexible classical environment~\cite{jiang2016nuclear}. Indeed, after training the parameters of $\{\epsilon_0, \gamma\}\cup\{\epsilon_m\}$ with data associated with the time interval $t\in[0,70]$fs, DIQCD models agree well with the data on $\langle \hat{\sigma}_x(t) \rangle$ and $\langle \hat{\sigma}_y(t)\rangle$ (see  Fig.~\ref{fig:rubrene}(b) for the case of $T=300$K as an example).

The one-site model is generalized for the lattice as 
\begin{equation}
\small
    \hat{H}_{\eps}(t)= \hat{H}_{\text{e}} + \sum_n \sum_{m=1}^9 \epsilon^{(n)}_m(t)c^\dagger_n c_n
\end{equation}
with the jump operator $\hat{L}_n=c^\dagger_n c_n$ (damping rate is $\gamma$) on any site-$n$. We estimate the carrier mobility by simulating Eq.~\eqref{lindblad} with the initial condition that one carrier is localized on the middle site of a finite molecular lattice of size $L=150$ and intermolecular distance $D=7\mathrm{\AA}$. The diffusion of the carrier at $T=300K$ is illustrated by Fig.~\ref{fig:rubrene}(d-e), showing a long-term behavior consistent with classical Fick's law. The mobility is calculated as $\mu(T)=\frac{e D^2}{2k_BT}\lim_{t\rightarrow \infty} \frac{d}{dt} \big\langle \tr(\rho_{\boldsymbol{\epsilon}}(t)n^2) -\tr^2(\rho_{\boldsymbol{\epsilon}}(t) n )\big\rangle_{\boldsymbol{\epsilon}}$. 
The estimation obtained from DIQCD agree closely with experimental estimation of intrinsic mobility~\cite{takeya2007very}, as well as the $\mu$ calculated from TD-DMRG~\cite{li2020finite} for a system with $L=21$ and $D=7\mathrm{\AA}$. 
The comparisons are shown in Fig.~\ref{fig:rubrene}(c), which also reports the $\mu$ we calculated from the Ehrenfest approach~\cite{wang2011mixed} (with non-adiabatic force). As expected, the Ehrenfest approach deviates more significantly from TD-DMRG and DIQCD at lower temperatures. It is unclear whether the minor but persistent discrepancy between DIQCD and TD-DMRG is due to the finite-size error in TD-DMRG or the approximation made by DIQCD. However, more importantly, DIQCD and TD-DMRG yield almost the same temperature-induced variation in $\mu$, indicating that DIQCD can mimic subtle variations in the quantum kinetic energy that outweighs the classical fluctuation in shaping the transport.

Our approach can be readily applied to a two-dimensional lattice model. Point defects can also be introduced by training several one-site DIQCD models that combine into an inhomogeneous lattice model. 
 
{\bf Discussion} -- 
DIQCD strikes a balance between accuracy and efficiency. In both case studies, DIQCD is trained on data from localized subsystems, reflecting typical real-world data collection. The resultant model is then generalized to the whole system by introducing known long-range interactions (dipolar interaction and intermolecular hopping). A key presumption here is that the noise in long-range interaction is either negligible or can be described with a known model, which holds for our two case studies. If this assumption does not hold, DIQCD would require training on data from a larger portion of the system to capture correlated noise. This may require approaches like TD-DMRG or multi-configuration time-dependent Hartree~\cite{meyer1990multi} if the data source is numerical simulations. 

Although explicit semiclassical approximation are avoided, DIQCD requires a clear understanding of the quantum device and the observables that govern the dynamical process relevant to device metrics. This challenge is similar to selecting a ``good'' set of collective variables for molecular conformational dynamics. Although physical intuitions are available for the case studies presented here, for general open quantum systems, one needs a systematic method for defining a canonical transformation that reveals the most relevant low-dimensional observables for data collection. 
 
 
{\bf Data and Code Availability} -- 
The Python package QEpsilon is publicly available at Github~\cite{qepsilon}. A tutorial for using QEpsilon is provided in the online documentation~\cite{qepsilon_doc}. All results reported in this paper can be reproduced through annotated code examples (``QEpsilon$\slash$examples$\slash$CaF$\_$qubits''  
and ``Qepsilon$\slash$examples$\slash$Rubrene$\_$Crystal'') within the package.

{\bf Acknowledgement} -- 
P.X. thanks Yukai Lu for his tremendous help with the specifications of the CaF qubit experiments and his contributions to preparing the manuscript. P.X. thanks Weitang Li for his help with the specifications of the Holstein model of Rubrene. We thank Zhen Huang, Lin Lin and Roberto Car for fruitful discussion. 
P.X. was supported by the Alvarez Fellowship of Lawrence Berkeley National Lab under contract No. DE-AC02-05CH11231.  
K.W. and X.L. were supported by the NSF Grants No. DMS-2111221 and No. CCF-2312456. 
C.Y. and Y.Z. were supported by the U.S. Department of Energy, Office of Science, Accelerated Research in Quantum Computing Centers, Quantum Utility through Advanced Computational Quantum Algorithms, grant no. DE-SC0025572. A. M. and W.A.dJ. were supported  by the U.S. Department of Energy under Contract No. DE-
AC02-05CH11231 Office of Science, Accelerated Research in Quantum Computing Centers, FAR-QC. 
Calculations were performed on the National Energy Research Scientific Computing Center (NERSC), a U.S. Department of Energy Office of Science User Facility operated under Contract No. DE-AC02-05CH11231.

\bibliography{apssamp}

\onecolumngrid
\section*{Appendices}
\twocolumngrid

\textbf{Appendix A - QEpsilon}
\begin{figure*}
    \centering
    \includegraphics[width=0.7\linewidth]{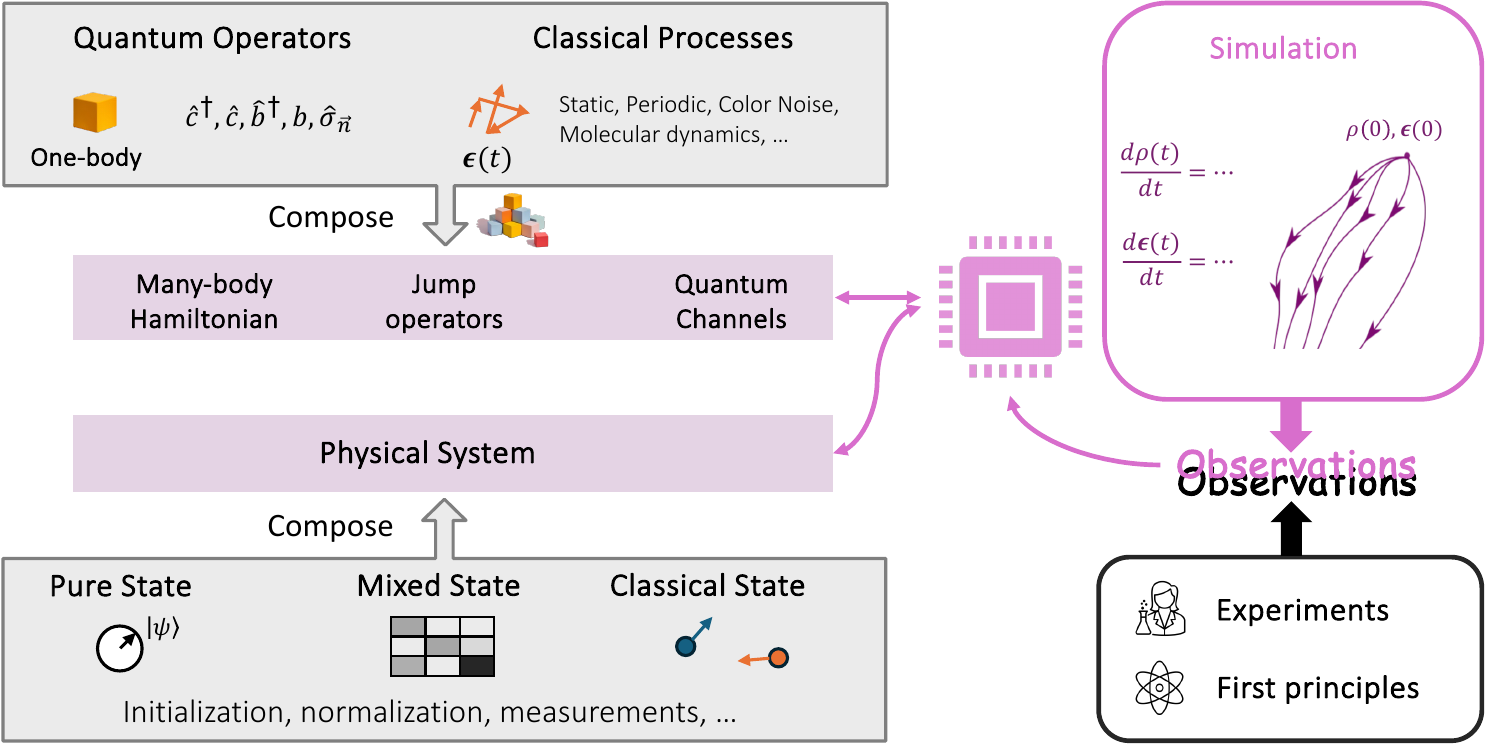}
    \caption{ The structure of QEpsilon.} 
    \label{fig:qepsilon}
\end{figure*}

A schematic representation of the structure of QEpsilon is shown in Fig.~\ref{fig:qepsilon}. 
QEpsilon can be roughly divided into the following modules: (1) a library of second-quantized bosonic, spin, and tight-binding operators as the basis for composing a many-body Hamiltonian and the jump operators; (2) a library of typical classical-process integrators for parameterizing $\eps(t)$; (3) a user interface for composing a many-body mixed system with these libraries; (4) structure-preserving integrators for simulating unitary and Lindblad dynamics (we adopt the $M=1$ integrator in Ref.~\cite{cao2021structure} for Lindblad equation); (5) a mechanism to train a DIQCD model on time-series data with backpropogation. 
These modularized components allow training and large-scale simulation in a unified, GPU-enabled framework. More technical details of QEpsilon should be found in the online documentation of the open-source package~\cite{qepsilon}. 
 
\textbf{Appendix B: CaF molecular qubits}

{\it Optical Tweezer} -- 
A single optical tweezer has a Guassian trapping potential $U_{\text{T}}(r,z) = -V\left(\frac{w_0}{w(z)}\right)^2 e^{\frac{-2r^2}{w^2(z)}}$
with respect to radial distance $r$ and height $z$ from the center. $V_{\text{m}}=k_B \times 1.28$mK is the maximal depth of the trap. $w_0=730$nm is the waist of the beam while $w(z)=w_0\sqrt{1+(z/z_R)^2}$ with the Rayleight range $z_R=\pi w^2_0 / \lambda$. $\lambda=781$nm is the wavelength. When performing a two-qubit gate, the central depth of the trap is limited to $V\approx k_B \times 0.13$mK.

{\it XY8 sequence} -- The size of each XY8 block (see Fig.~\ref{fig:caf}) is 1.6ms for one-qubit experiments and two-qubit experiments with $d< 1.68\mathrm{\mu m}$, 3.2ms for two-qubit experiments with $d\geq 1.68 \mathrm{\mu m}$. 

{\it CaF molecules} -- The coupling constant for dipole-dipole interaction is $J_0= 1942.3 \mathrm{\hbar \cdot Hz \cdot \mu m^3}$~\cite{holland2023demand}.
When performing the two-qubit gate, Raman thermometry gives a rough estimation of the molecular temperature that amounts to $T_r=6\mathrm{\mu K}$ for radial motion (parallel to $xy$ plane) and $T_a=18\mathrm{\mu K}$ for axial motion ($z$ direction). Note that the molecular temperature changes with the depth of the optical trap. So the temperature here is lower than the temperature estimated at full depth~\cite{holland2023demand}.

{\it Bell State creation} -- The scheme is illustrated by Fig.~\ref{fig:caf}(b). The initial state is $\ket{\uparrow\uparrow}$. Two $\pi/2$ pulses along the $x$ axis are applied at the start and the end of the scheme. $XY8$ sequences are applied consecutively for error suppression.

{\it Molecular dynamics of $R(t)$} -- A Langevin integrator ~\cite{zhang2019unified} is used to simulate the molecular dynamics of $R(t)$ and its velocity $v(t)$, given as
\begin{equation}
\small
\begin{cases}
    v^{\text{mid}}_i(t) &= v_i(t) + F_i(t) \delta t / m \\
    R^{\text{mid}}_i(t) &= R_i(t) + 0.5 v^{\text{mid}}_i(t) \delta t \\
    v_i(t+\delta t) &= e^{-\delta t/\tau} v^{\text{mid}}_i(t) + A_i \sqrt{1-e^{-2\delta t/\tau}} \xi_i(t) \\
    R_i(t+\delta t) &= R_i(t) + 0.5 v_i(t+\delta t) dt
\end{cases}
\end{equation}
$F_i(t)$ is the gradient force from the optical trap on the $i$-th ($i\in [1,6]$) component of $R$. 
The radial ($x$ and $y$ directions) and axial ($z$ direction) motion are controlled ~\cite{kubo1966fluctuation} with the same damping time ($\tau=100$ms) but different amplitude ($A$) of white noise ($\xi$) to maintain their respective temperature $T_r=6\mathrm{\mu K}$ and $T_a=18\mathrm{\mu K}$.

\begin{table}
    \begin{ruledtabular}
    \begin{tabular}{cccccccccc}
        $m$ & 1 & 2 & 3 & 4 & 5 & 6 & 7 & 8 & 9\\
        $\omega_m / \mathrm{cm^{-1}}$  & 84  & 214  &632  & 1002  & 1206 & 1351 & 1364 & 1535 & 1594 \\
        $g_m$ & 0.96 & 0.37 & 0.25 & 0.20& 0.15 & 0.31 & 0.13 & 0.20 & 0.31 \\
    \end{tabular}
    \end{ruledtabular}
    \caption{Frequency of kept vibrational modes and the renormalized coupling constants.}
    \label{tab:couple}
\end{table}

\begin{figure*}
    \centering
    \includegraphics[width=\linewidth]{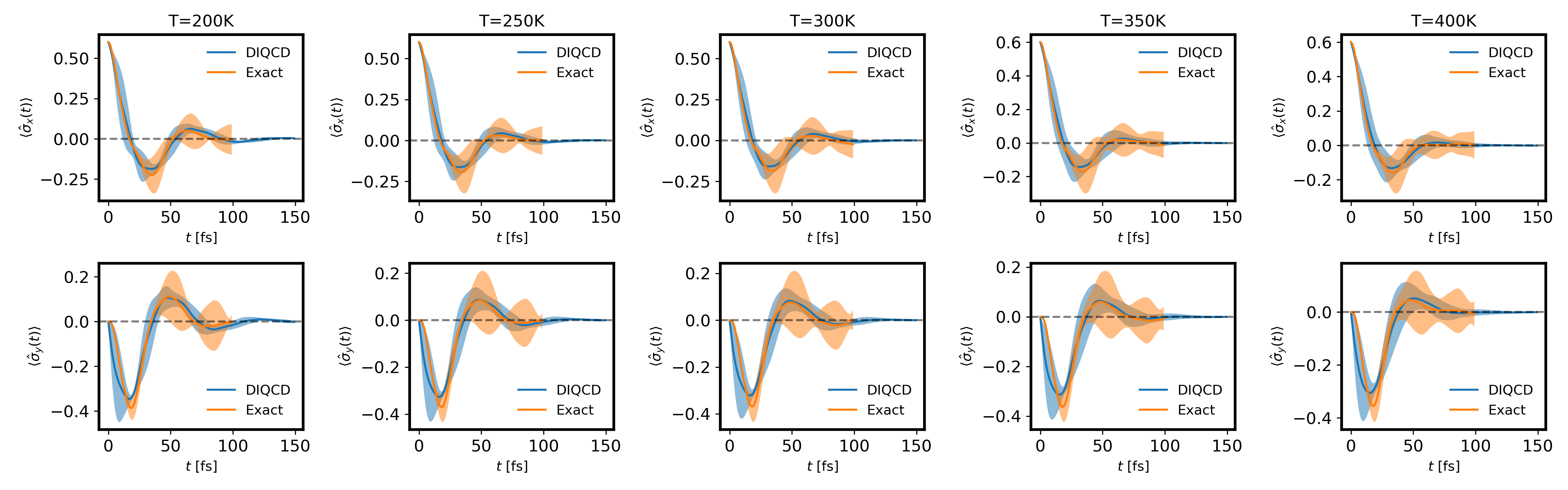}
    \caption{  The mean and standard deviation (concerning the ensemble of sampled trajectories) of $\langle\hat{\sigma}_x(t)\rangle $ and $\langle\hat{\sigma}_y(t)\rangle$ from the final DIQCD models and the data associated with exact unitary simulation. } 
    \label{fig:train_rubrene}
\end{figure*}

{\it Training} -- We train the DIQCD model with all 24 one-qubit data (see Fig.~\ref{fig:caf}) simultaneously. For each epoch, we simulate DIQCD three times, respectively with plain, spin-echo, and XY8 control schemes. The batch size for each simulation is 512. The time step for integration is 0.01ms for the plain scheme, 0.05ms for the spin-echo scheme, and 0.1ms for the XY8 scheme. The loss function is $L=\frac{1}{7}\sum_{i=1}^{7} (C^{\text{M,Plain}}_i - \varsigma^{-1}C^{\text{D,Plain}}_i)^2 + \frac{1}{10}\sum_{i=1}^{10}(C^{\text{M,Echo}}_i - \varsigma^{-1}C^{\text{D,Echo}}_i)^2 + \frac{1}{7}\sum_{i=1}^{7} (C^{\text{M,XY8}}_i - \varsigma^{-1}C^{\text{D,XY8}}_i)^2$. The superscript ``M'' and ``D'' indicate the Model and the Data, respectively. $C^{\text{D,Plain}}_i$ is the $i$-th ($i=1,2,\cdots,7$) data point associated with the plain scheme. $C^{\text{M,Plain}}_i$ is the corresponding DIQCD model prediction. The meaning of $C^{\text{M,Echo}}_i$, $C^{\text{D,Echo}}_i$, $C^{\text{M,XY8}}_i$, $C^{\text{D,XY8}}_i$ are self-evident.  The DIQCD model is trained for 200 epochs, with the standard ADAM optimizer in PyTorch with a fixed learning rate of 0.1 for all parameters but the damping rates $\gamma_x$ and $\gamma_z$. The learning rate for damping rates is set to 0.001. The loss saturates after roughly 50 epochs. The final productive DIQCD model is chosen to be the model trained after all the epochs. The training is done with QEpsilon and a CPU node on Perlmutter. Training with a GPU is not necessary for a small quantum system.  

\textbf{Appendix C: Rubrene Crystal}

{\it Electron-phonon coupling} -- Parameters for electron-phonon coupling in Rubrene were derived in Ref.~\cite{jiang2016nuclear, li2020finite} from density functional theory calculations with B3LYP functional approximation~\cite{Becke1993}, the 6-31G(d) basis set, and the Gaussian 09 software~\cite{g09}. 
Each Rubrene molecule has 210 vibrational modes, with tens of them contributing to electron-phonon coupling. To reduce the complexity of the effective Hamiltonian describing electron-phonon interaction, one can keep only a few modes with significant electron-phonon coupling and renormalize their coupling constants to conserve the total reorganization energy. The effective model adopted here keeps nine modes, with frequency ($\omega_m$) and coupling constant ($g_m$) listed in Table~\ref{tab:couple}.

{\it Data generation} -- We generate our data by simulating the plain unitary quantum dynamics with a timestep of 0.01fs. We use $\phi=0.1$ throughout the work. The total sample time for each unitary simulation is 100fs. For each temperature (T=200K,250K,300K,350K,400K) separately, we generate 64 trajectories of unitary evolution. Then, we calculate the mean and standard deviation on observing $\hat{\sigma}_x(t)$ and $\hat{\sigma}_y(t)$ at $t=0,1,\cdots, 99$fs. The simulation is done with QEpsilon and one NVIDIA A100 GPU. 

{\it Training} -- For each temperature separately, we train a DIQCD model with a subset of one-molecule data associated with the sampling time interval $t\in[0,70]$fs.  The loss function is $L=\sum_{i=1}^{70} (\mathrm{Mean^{M}}[\hat{\sigma}_x(i \cdot \text{fs})] -\mathrm{Mean^D}[\hat{\sigma}_x(i \cdot \text{fs})])^2 + (\mathrm{MEAN^{M}}[\hat{\sigma}_y(i \cdot \text{fs})] -\mathrm{MEAN^D}[\hat{\sigma}_y(i \cdot \text{fs})])^2 + 0.1(\mathrm{STD^{M}}[\hat{\sigma}_x(i \cdot \text{fs})] -\mathrm{STD^D}[\hat{\sigma}_x(i \cdot \text{fs})])^2 + 0.1(\mathrm{STD^{M}}[\hat{\sigma}_y(i \cdot \text{fs})] -\mathrm{STD^D}[\hat{\sigma}_y(i \cdot \text{fs})])^2$. The superscript ``M'' and ``D'' indicate the Model and the Data, respectively. ``MEAN'' and ``STD'' represent mean value and standard deviation concerning the 64 trajectories. Note that the loss concerning standard deviation weights 0.1 due to its minor regularization role. 
The forward evolution of DIQCD uses a batch size of 512 and a timestep of 0.05fs.  Each DIQCD model is trained for 200 epochs of forward evolution, with the standard ADAM optimizer in PyTorch with a fixed learning rate of 0.3.  Model parameters converge after roughly 150 epochs. The final productive DIQCD model is chosen to be the model trained after all the epochs. 
The training is done with QEpsilon and one NVIDIA A100 GPU. A comparison between the final DIQCD models and the data is provided in Fig.~\ref{fig:train_rubrene}.

{\it Simulation} -- Simulation of the DIQCD models for a molecular lattice of size $L=150$ is done with QEpsilon and one NVIDIA A100 GPU. The batch size for forward evolution is 512, and the timestep is 0.01fs. The total sample time is 100fs for calculating the carrier mobility.

\end{document}